\documentclass[prl,noshowpacs,english,superscriptaddress,twocolumn]{revtex4}
\usepackage{amsmath,amssymb,amsfonts,mathrsfs}
\usepackage{amsthm}
\usepackage{CJK}
\usepackage{bm}
\usepackage{babel}
\usepackage{graphicx}
\allowdisplaybreaks
\usepackage{braket}
\usepackage[breaklinks]{hyperref}
\hypersetup{colorlinks=true, linkcolor=blue, citecolor=blue, filecolor=blue, urlcolor=blue}

\begin{document}

\title{Three-Dimensional Quantum Anomalous Hall Effect in  Ferromagnetic Insulators}

\author{Y. J. Jin}
\affiliation{Department of Physics $\&$ Institute for Quantum Science and Engineering, Southern University of Science and Technology, Shenzhen 518055, P. R. China.}
\author{R. Wang}
\affiliation{Department of Physics $\&$ Institute for Quantum Science and Engineering, Southern University of Science and Technology, Shenzhen 518055, P. R. China.}
\affiliation{Institute for Structure and Function $\&$
Department of physics, Chongqing University, Chongqing 400044, P. R. China.}
\author{B. W. Xia}
\affiliation{Department of Physics $\&$ Institute for Quantum Science and Engineering, Southern University of Science and Technology, Shenzhen 518055, P. R. China.}
\author{B. B. Zheng}
\affiliation{Department of Physics $\&$ Institute for Quantum Science and Engineering, Southern University of Science and Technology, Shenzhen 518055, P. R. China.}
\author{H. Xu}
\email[]{xuh@sustc.edu.cn}
\affiliation{Department of Physics $\&$ Institute for Quantum Science and Engineering, Southern University of Science and Technology, Shenzhen 518055, P. R. China.}

\begin{abstract}

The quantum anomalous Hall effect (QAHE) hosts the dissipationless chiral edge states associated with the nonzero Chern number, providing potentially significant applications in future spintronics. The QAHE usually occurs in a two-dimensional (2D) system with time-reversal symmetry breaking. In this work, we propose that the QAHE can exist in three-dimensional (3D) ferromagnetic insulators. By imposing inversion symmetry, we develop the topological constraints dictating the appearance of 3D QAHE based on the parity analysis at the time-reversal invariant points in reciprocal space. Moreover, using first-principles calculations, we identify that 3D QAHE can be realized in a family of intrinsic ferromagnetic insulating oxides, including layered and non-layered compounds that share a centrosymmetric structure with space group $R\bar{3}m$ (No. 166). The Hall conductivity is quantized to be $-\frac{3e^2}{hc}$ with the lattice constant $c$ along $c$-axis. The chiral surface sheet states are clearly visible and uniquely distributed on the surfaces that are parallel to the magnetic moment. Our findings open a promising pathway to realize the QAHE in 3D ferromagnetic insulators.

\end{abstract}


%Use showkeys class option if keyword %display desired

\maketitle
A two-dimensional (2D) electron gas exhibits the integer quantum Hall effect (IQHE)\cite{Tsui1982} at low temperatures in strong magnetic fields, which offers an alternative avenue to achieve a dissipationless current beyond superconductors.
The IQHE has one-dimensional (1D) chiral edge states originated from the 2D bulk band topology, giving rise to the quantized Hall conductance in the units of $e^2/h$.
The topological properties of the IQHE are characterized by a topological invariant $\mathcal{C}$ known as the first Chern number \cite{PhysRevLett.49.405}. However, the applications of IQHE are strongly limited by the requirement of the high-intensity external magnetic field and low temperature. The most promising solution to these challenges is to realize the quantum anomalous Hall effect (QAHE) in insulating materials co-existing with the band topology and ferromagnetic (FM) order \cite{Haldane1988,Yu2010,PhysRevB.85.045445,Chang2013,Fang2014,Chang2015,PhysRevLett.115.186802}.
The QAHE insulators (i.e., Chern insulators) host nontrivial topological properties associated with the nonzero Chern number, and provide significant applications in topological spintronics. Hence, extensive investigations have always been performed to propose candidate materials to realize such QAHE. Up to now, there are numerous theoretically predicted candidates to possess the QAHE, such as intrinsic or doped magnetic topological insulators \cite{Yu2010, PhysRevLett.115.186802,PhysRevLett.117.056804,PhysRevLett.119.026402,PhysRevB.95.201402,PhysRevB.95.134448,PhysRevB.95.125430, PhysRevB.95.045113}. Recently, the QAHE was observed experimentally in Cr-doped (Bi,Sb)$_2$Te$_3$ thin films at a low temperature around 30 mK \cite{Chang2013}. Unfortunately, the QAHE has not been realized in intrinsic FM materials due to the lack of suitable materials. Alternatively, it is promising to investigate the exotic properties of QAHE in three-dimensional (3D) QAHE insulators since the 3D materials with long-range magnetic order host higher thermodynamic stability than 2D ones.

\begin{figure}
	\centering
\setlength{\belowcaptionskip}{-0.35cm}
\setlength{\abovecaptionskip}{-0.05cm}
	\includegraphics[scale=0.29]{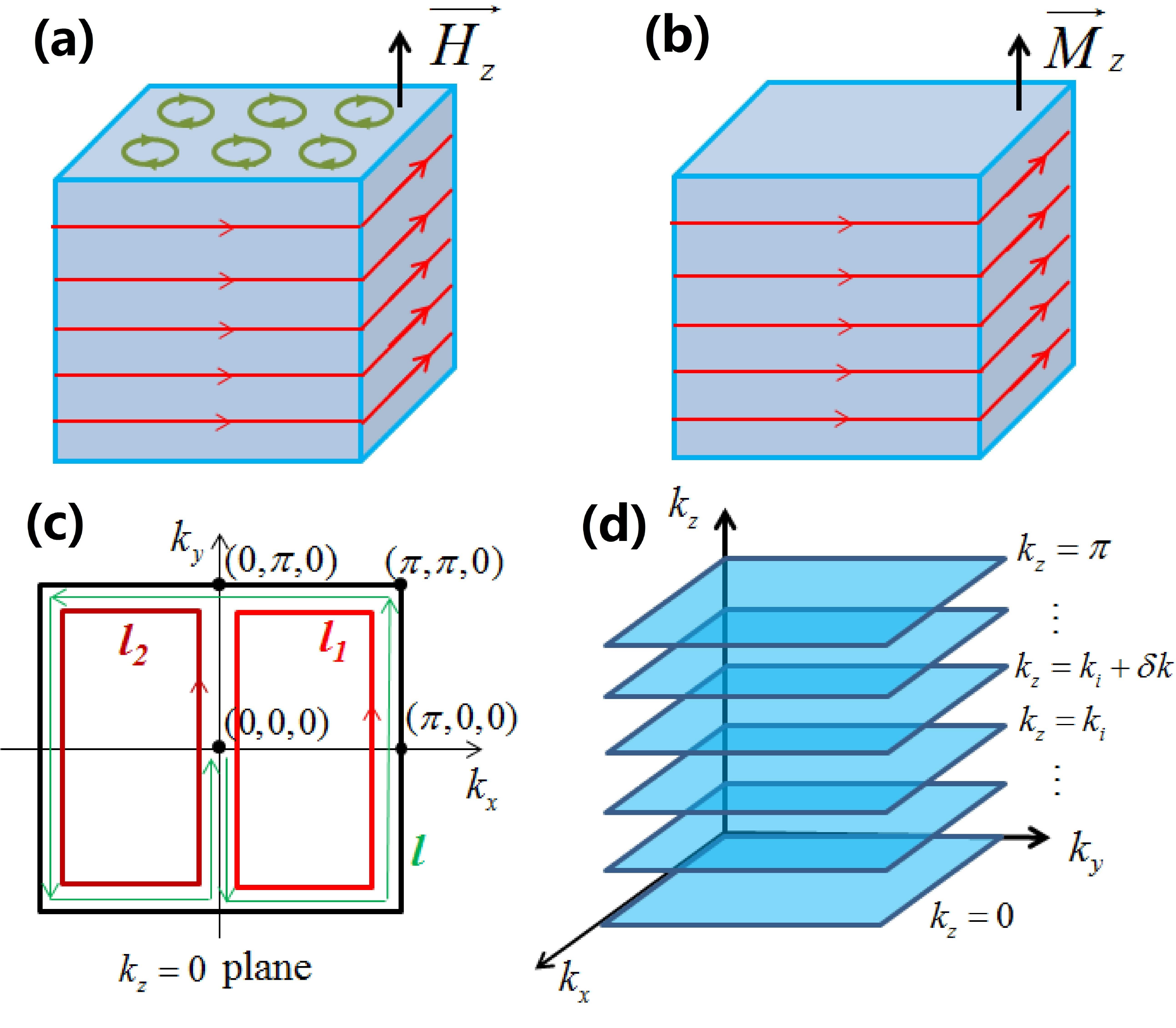}
	\caption{(a) In a strong magnetic field $\mathbf{H}_z$, the 3D electron gas exhibits the 3D IQHE with the chiral surface sheet states due to the zero modes without $k_z$ dispersion. (b) The 3D QAHE in a FM insulator with a intrinsic magnetic moment $\mathbf{M}_z$ with the chiral surface sheet states. (c) The invariant plane with $k_z =0$ in a cubic BZ. The paths $l$, $l_1$, and $l_2$ are used in Berry phase arguments. The TRIM points in this plane are denoted. (d) The adiabatic interpolation between the planes $k_z=0$ and $k_z=\pi$. Each imaginary 2D cut orthogonal to the $k_z$ axis is associated with a fixed Chern number.
\label{figure-chial}}
\end{figure}

To illustrate the 3D QAHE, we first recall a fact of a 3D electron gas system in a strong magnetic field, which was first studied by Halperin \cite{Halperin1987} in 1987. When the Fermi level lies inside an energy gap, the 3D IQHE may emerge. The Hall conductivity tensor is quantized and given by the Kohmoto-Halperin-Wu formula as $\sigma_{ij}=\frac{e^2}{2\pi h} \epsilon_{ijk}G_{k}$ \cite{Halperin1987,PhysRevB.45.13488}, where $\epsilon_{ijk}$ is the fully antisymmetric tensor and $\mathbf{G}$ is a reciprocal-lattice vector. The 3D IQHE exhibits chiral surface sheet states due to the appearance of an energy spectrum without $k_z$ dispersion under a strong magnetic field $\mathbf{H}_z$ [see Fig. \ref{figure-chial}(a)] \cite{PhysRevB.45.13488}. Even through the presence of 3D IQHE requires more stringent conditions, this phenomenon has been predicted in 3D Hosftadter butterfly's system \cite{PhysRevLett.86.1062}, graphite \cite{PhysRevLett.99.146804}, and 3D topological semimetals \cite{PhysRevLett.119.136806}, etc. Analogously, if chiral surface sheet states occur on a specific surface of a 3D FM insulator with intrinsic magnetic moment $\mathbf{M}_z$, we can define the 3D QAHE [see Fig. \ref{figure-chial}(b)]. The 3D QAHE insulator can be viewed as a series of imaginary 2D QAHE insulators in momentum space, for which the Chern number of each $k_z$ plane remains unchanged in momentum space \cite{PhysRevB.83.245132,PhysRevB.95.075146}. However, the Chern number may generally vary as a function of $k_z$ and the bulk gap is closed, resulting in a 3D topological semimetals \cite{Xu2011,PhysRevLett.107.127205}. In Weyl semimetals, a 3D QAHE phase as a 2D subsystem may appear, which does not possess a global band gap \cite{PhysRevB.83.205101,PhysRevB.97.201101}. So far, it is challenging to explore 3D QAHE insulators with fully nontrivial bulk gaps.

First of all, we elucidate the topological constraints in FM compounds with inversion ($\mathcal{I}$) symmetry. In a 3D system with breaking time-reversal ($\mathcal{T}$) but keeping $\mathcal{I}$ symmetry , we introduce that the topological invariant of characterizing a 3D QAHE insulator can be determined by the parity eigenvalues of occupied states at eight time-reversal invariant momenta (TRIM) points. A 3D FM insulating Hamiltonian $\mathcal{H}(\mathbf{k})$ that is invariant under $\mathcal{I}$ symmetry is considered. For convenience, we take the $k_z =0$ plane of a cubic Brillouin zone (BZ). The 2D Hamiltonian $\mathcal{H}(\mathbf{k})|_{k_z=0}$ also possesses the $\mathcal{I}$ symmetry [see Fig. \ref{figure-chial}(c)]. The 2D Hall conductance in $k_z =0$ plane is expressed as \cite{Haldane1988}
\begin{equation}
\sigma_{xy}^{\mathrm{2D}}|_{k_z=0}=\frac{e^2}{2\pi h}\oint_{l}\mathbf{A}(\mathbf{k})\cdot d\mathbf{k}=-\frac{e^2}{h}{\mathcal{C}},
\end{equation}
where $\mathbf{A}(\mathbf{k})=-i\sum_{i}{\langle u_{n}(\mathbf{k})|\nabla_{\mathbf{k}}|u_{n}(\mathbf{k})\rangle}$ is Berry connection characterized by the occupied Bloch eigentstates $u_n(\mathbf{k})$. The 2D BZ boundary $l$ can be viewed as the composition $l_1+l_2$, and then the Berry phase $\oint_{l}\mathbf{A}(\mathbf{k})\cdot d\mathbf{k}=\oint_{l_1}\mathbf{A}(\mathbf{k})\cdot d\mathbf{k}+\oint_{l_2}\mathbf{A}(\mathbf{k})\cdot d\mathbf{k}$. The $\mathcal{I}$ symmetry  guarantees that $\oint_{l_1}\mathbf{A}(\mathbf{k})\cdot d\mathbf{k}=\oint_{l_2}\mathbf{A}(\mathbf{k})\cdot d\mathbf{k}$, and we can obtain
\begin{equation}\label{chernb}
(-1)^{\mathcal{C}|_{k_z=0}}=e^{i\oint_{l_1}\mathbf{A}(\mathbf{k})\cdot d\mathbf{k}}=\prod_{\mathbf{K}_{\mathrm{inv}}}\xi(\mathbf{K}_{\mathrm{inv}})|_{k_z=0},
\end{equation}
where $\xi(\mathbf{K}_{\mathrm{inv}})=\prod_{n}\xi_{n}(\mathbf{K}_{\mathrm{\mathrm{inv}}})$ and $\xi_{n}(\mathbf{K}_{\mathrm{inv}})$ is the parity of the occupied Bloch states $u_n(\mathbf{K}_{\mathrm{inv}})$ at the TRIM point $\mathbf{K}_{\mathrm{inv}}$. Eq. (\ref{chernb}) indicates that the first Chern number $\mathcal{C}$ connects the Berry phase of half BZ in $k_z=0$ plane and parity eigenvalues at the TRIM points. The detailed proof is provided in the Supplemental Material (SM) \cite{SM}.
Similarly, the Chern number in the other invariant plane is the same.
%\begin{equation}\label{chernb1}
%(-1)^{\mathcal{C}|_{k_z=\pi}}=\prod_{\mathbf{K}_{\mathrm{inv}}}\xi(\mathbf{K}_{\mathrm{inv}})|_{k_z=\pi}.
%\end{equation}

Due to the assumption of the Hamiltonian describing a 3D FM insulator, the adiabatic interpolation between the planes $k_z=0$ and $k_z=\pi$ can be thought as a series of imaginary 2D FM insulator with the fixed Chern number [see Fig. \ref{figure-chial}(d)].  The Chern number of all the 2D cuts at a certain $k_z$ is the same since those planes at two momenta $k_z=k_i$ and $k_z=k_{i}+\delta k$ can be adiabatically connected without closing the gap. Then, we can obtain $\mathcal{C}|_{k_z=0}=\mathcal{C}|_{k_z=\pi}$. Hence, for a 3D FM insulator, if any 2D cut with first Chern number $\mathcal{C}=1$ is identified, this material must be a 3D QAHE insulator. Based on the discussions above, we propose a general recipe to identify a 3D QAHE insulator using the parity eigenvalues: (1) a 2D cut with four TRIM points is a 2D Chern insulator; (2) the product over the inversion eigenvalues of all occupied bands at eight TRIM points must be $1$.

This connection to the parity eigenvalues allows us to explore or design 3D QAHE insulators. We can perform high-throughput calculations to identify the band topology based on the parity eigenvalues at the TRIM points. Generally, such a route can be taken when simulating FM layered materials with $\mathcal{I}$ symmetry, in which the interlayer hopping is weak. However, our topological constraints don't require that a 3D QAHE insulator must be formed as a stacking of 2D QAHE insulators in real space. In that sense, 3D QAHE insulators can not only be present in layered ferromagnetic insulators, but also emerge in non-layered ones. As expected, we show that the 3D QAHE is realized in a family of FM insulating oxides with space group $R\bar{3}m$ (No. 166), including layered Ba$_{2}$Cr$_{7}$O$_{14}$, Sr$_2$Cr$_7$O$_{14}$, K$_2$V$_7$O$_{14}$ and non-layered hexagonal ($h$-)Fe$_3$O$_4$. In the main text, we mainly focus on a layered Ba$_{2}$Cr$_{7}$O$_{14}$, which has already been synthesized in 1972 \cite{Evans1972}. The brief results of the non-layered $h$-Fe$_3$O$_4$ is also provided to illustrate the topologically nontrivial features of 3D QAHE. Detailed results of $h$-Fe$_3$O$_4$ and other layered compounds in the family are included in the SM \cite{SM}.

To verify the existence of 3D QAHE in the family of FM oxides, we perform first-principles calculations using the Vienna \textit{ab initio} Simulation Package \cite{Kresse2} based on density functional theory \cite{Kohn}. The Perdew-Burke-Ernzerhof (PBE)-type generalized gradient approximation (GGA) \cite{Perdew1,Perdew2} is chosen for the exchange-correlation potential. The core-valence interactions are treated by the projector augmented wave (PAW) method \cite{Kresse4,Ceperley1980} with a plane-wave-basis cutoff of 600 eV. The full BZ is sampled by $15\times15\times15$ Monkhorst-Pack grid \cite{Monkhorst} to simulate the electronic behaviors. Since the $3d$-electrons of transition metals, i.e., Cr, Fe,  are correlated, we employ GGA+$U$ scheme \cite{Liechtenstein1995} and introduce the on-site Coulomb repulsion of $U = 3.0$ eV for Cr \cite{PhysRevLett.80.4305} and $U = 3.6$ eV for Fe \cite{PhysRevLett.99.206402}. Furthermore, we also confirm that the topological features are robust in a wide range of $U$.

\begin{figure}
\setlength{\belowcaptionskip}{-0.35cm}
\setlength{\abovecaptionskip}{-0.05cm}
	\centering
	\includegraphics[scale=0.24]{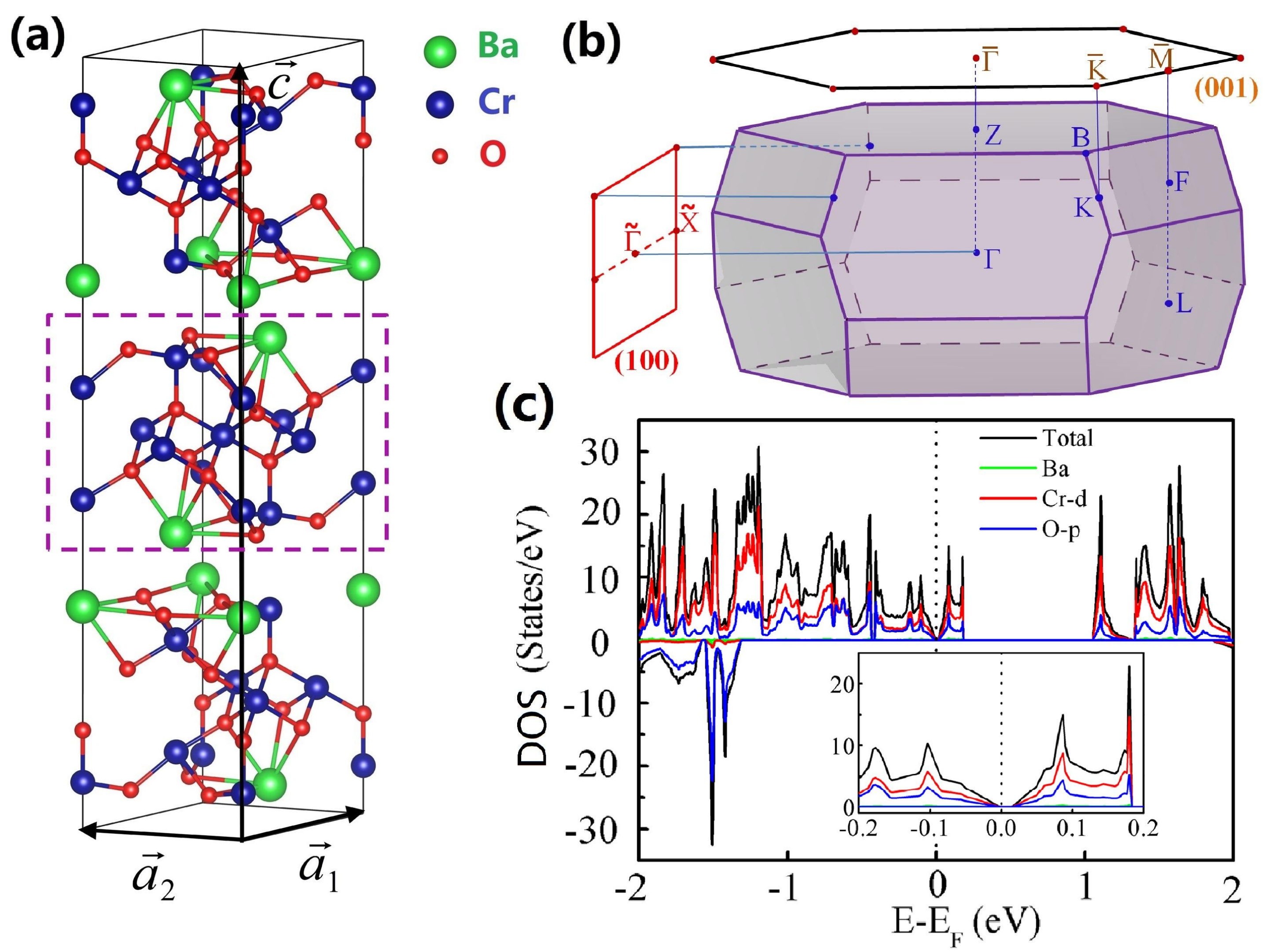}
    \includegraphics[scale=0.46]{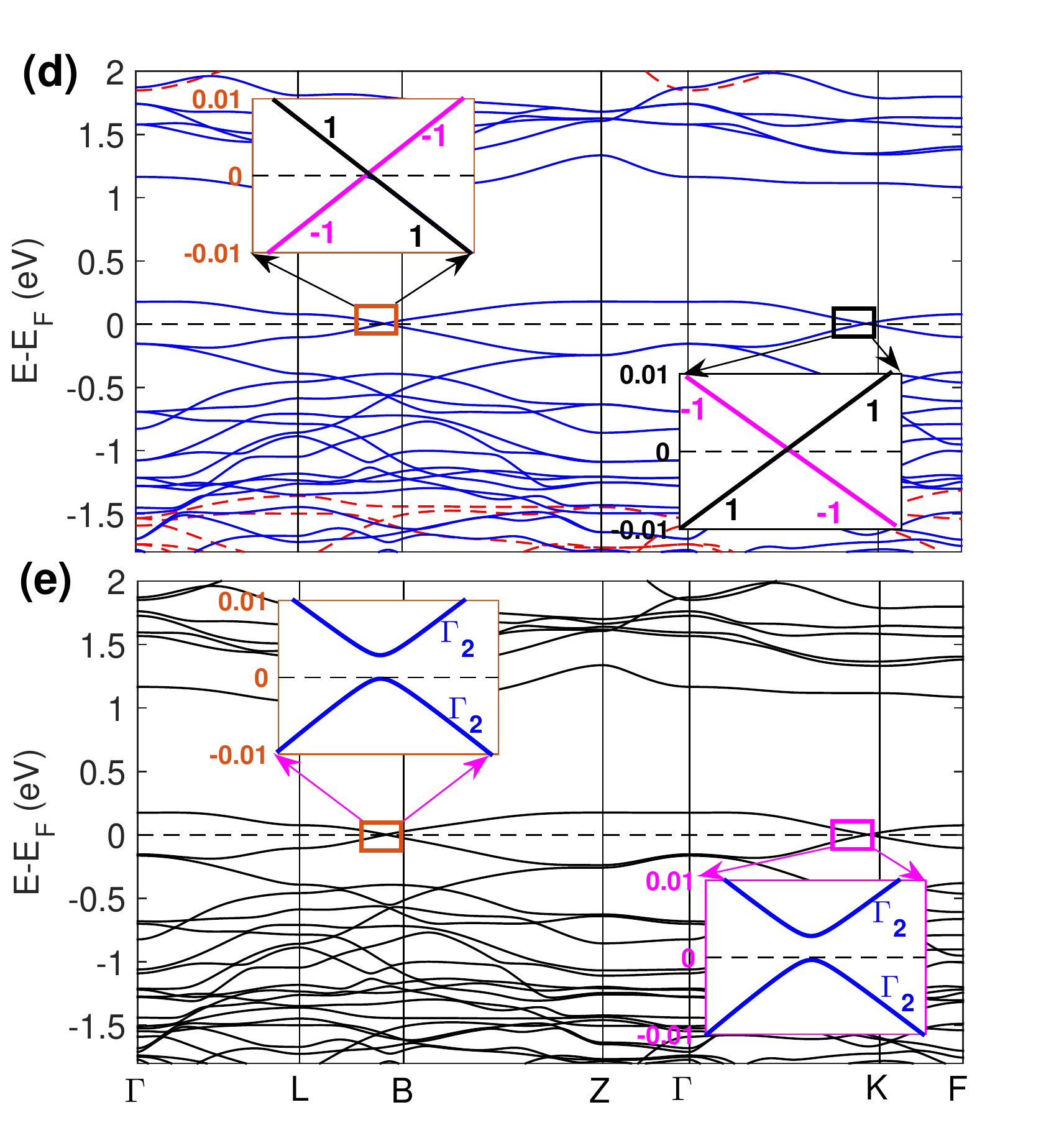}
	\caption{The crystal and band structures of Ba$_{2}$Cr$_{7}$O$_{14}$. (a) The unit cell and a single layer denoted by the dashed-boxed-region. (b) The 3D BZ and the corresponding (001) and (100) surface BZ. (c) Spin resolved DOS and partial DOS without SOC. (d) Band structure of  Ba$_{2}$Cr$_{2}$O$_{2}$ without SOC. The majority and minority spin bands are denoted by solid blue and dashed red lines, respectively. The two bands with opposite mirror eigenvalues $\pm$1 cross at the Fermi level in the $L$-$B$ and $\Gamma$-$K$ directions are shown as insets. (e) Band structure with SOC along [001] magnetization.
\label{struct}}
\end{figure}

We first show 3D QAHE in the layered compound Ba$_{2}$Cr$_{7}$O$_{14}$. It crystallizes in a rhombohedral structure with space group $R\bar{3}m$ (No. 166). As shown in Fig. \ref{struct}(a), it has a layered structure. Each layer possesses a hexagonal lattice with unit cell vectors $\mathbf{a}_1$ and $\mathbf{a}_2$, and $|\mathbf{a}_1|=|\mathbf{a}_2|=a$. The stacking period of Ba$_{2}$Cr$_{7}$O$_{14}$ is characterized by the third lattice vector $\mathbf{c}$. The Ba-terminations of both sides of each layer induce a weak interlayer bonding, so the van der Walls-type interlayer interaction is present. The optimized lattice constants are $a=5.870$ {\AA} and $c=27.931$ {\AA}, which are very close to the experimental values $a=5.652$ {\AA} and $c=27.770$ \cite{Evans1972}. The 3D BZ and the corresponding (001) and (100) surface BZ of Ba$_{2}$Cr$_{7}$O$_{14}$ are shown in Fig. \ref{struct}(b).

The spin resolved density of states (DOS) and partial DOS without SOC are shown in Fig. \ref{struct}(c), and the magnetic moment per Cr atom is $\sim$2.57 $\mu_B$. We find that the states around the Fermi level are completely contributed by the majority spin states of Cr-$3d$ and O-$2p$ orbitals, while all orbitals of Ba atom keep far away from the Fermi level. Our results strongly suggest that the FM ground state of Ba$_{2}$Cr$_{7}$O$_{14}$ originates from the half-metallic characteristic of CrO$_2$ \cite{PhysRevLett.80.4305}.  The band structure without SOC is shown in Fig. \ref{struct}(d). We find that two majority spin bands exactly cross the Fermi level in high-symmetry $L$-$B$ and $\Gamma$-$K$ directions, respectively. With the spin-rotation symmetry, the two crossing bands belong to opposite mirror eigenvalues $\pm1$.

In the presence of SOC, the magnetic anisotropic calculations indicate that the ground state hosts a magnetization along the stacking direction (i.e., the [001] direction). Figure \ref{struct}(e) shows the corresponding SOC band structure, which is very similar to that without SOC due to the weak SOC strength of Cr and O elements. When the magnetization is along the [001] direction, the symmetry of this system is reduced to the magnetic group $D_{3d}(S_6)$ that contains inversion $\mathcal{I}$, three-fold rotation symmetry $C_{3z}$, and antiunitary mirror symmetry $\mathcal{T}M_{x}$, where mirror plane of $M_x$ is perpendicular to $x$ axis. Hence, a bulk band gap $\sim3.5$ meV is present as no any symmetry can protect the band crossings. Therefore, the compound Ba$_{2}$Cr$_{7}$O$_{14}$ is a layered FM insulator. Some another compounds with bigger SOC strength in the family possess larger gaps and can be found in SM \cite{SM}.

\begin{figure}
\setlength{\belowcaptionskip}{-0.35cm}
\setlength{\abovecaptionskip}{-0.05cm}
	\centering
	\includegraphics[scale=0.28]{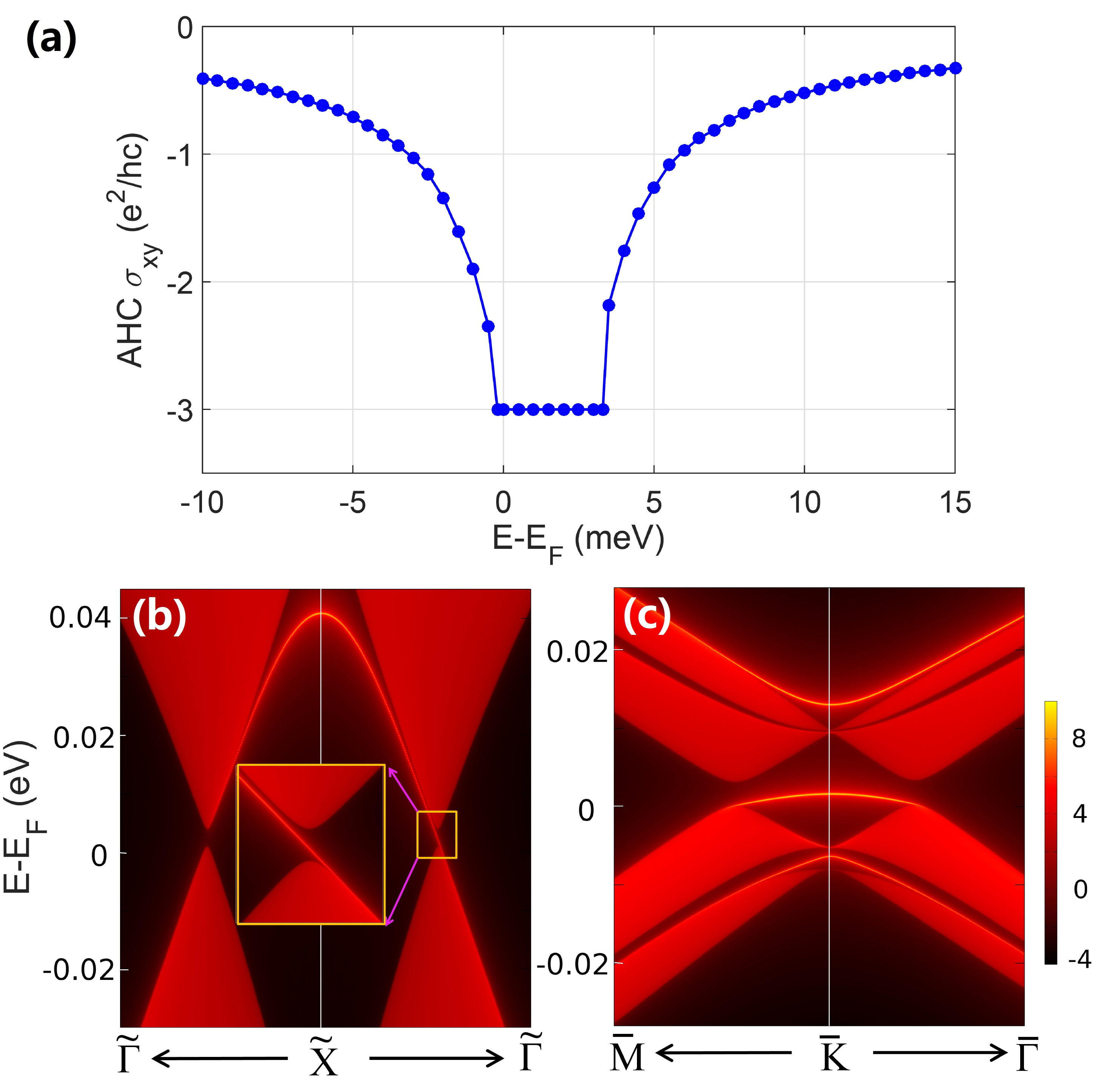}
	\caption{(a) The intrinsic Hall conductivity of Ba$_{2}$Cr$_{7}$O$_{14}$ relative to the Fermi level. The quantized plateau ($\sigma_{xy}^{\mathrm{3D}}=-3e^2/hc$) is shown. (b) The LDOS of Ba$_{2}$Cr$_{7}$O$_{14}$ projected on the (100) surface with the chiral surface sheet state. (c) The LDOS of Ba$_{2}$Cr$_{7}$O$_{14}$ projected on the (001) surface with trivial surface states.
\label{figure3}}
\end{figure}

Next, we use the parity analysis to demonstrate that Ba$_{2}$Cr$_{7}$O$_{14}$ is a 3D QAHE insulator. The parities of the occupied Bloch states at eight TRIM points are listed in Table \ref{tableI}. We can easily conclude that there are two paralleled planes with the first Chern number $\mathcal{C} = 1$ in reciprocal space. These two planes belong to a parallelepiped in which the eight TRIM points are the vertices. The one passes through $\Gamma$ point and the other passes through $Z$ point. Generally, the Chern number is invariant when the adiabatic deformation are performed in the manifold. Then, we can obtain that $\mathcal{C}=1$ for the planes $k_z=0$ and $k_z=3\pi /c $ under a smooth transformation, where $3\pi /c$ is the $z$-component of $Z$ point (0, 0, $3\pi /c$) in Cartesian coordinates. As a result, the insulating Ba$_{2}$Cr$_{7}$O$_{14}$ with the intrinsic magnetization along the [001] direction can exhibit the 3D QAHE. The 3D Hall conductivity can be obtained by the integral of $\sigma_{xy}^{\mathrm{2D}}$ as
\begin{equation}\label{hall3D}
\sigma_{xy}^{\mathrm{3D}}=\int_{-|\mathbf{G}_z|/2}^{|\mathbf{G}_z|/2}\frac{dk_z}{2\pi}\sigma_{xy}^{\mathrm{2D}}=-\frac{e^2}{2\pi h}\mathcal{C}|\mathbf{G}_z|,
\end{equation}
where $\mathbf{G}_z$ is the reciprocal lattice vector in the $z$ direction. Eq. (\ref{hall3D}) is consistent with the  Kohmoto-Halperin-Wu formula \cite{Halperin1987,PhysRevB.45.13488}. For Ba$_{2}$Cr$_{7}$O$_{14}$, $|\mathbf{G}_z|=6\pi /c$ (see SM \cite{SM}) and $\mathcal{C}=1$, i.e., $\sigma_{xy}^{\mathrm{3D}}=-3e^2/hc$. The intrinsic Hall conductivity of Ba$_{2}$Cr$_{7}$O$_{14}$ can also be calculated from the Kubo formula with the Wannier tight-binding (TB) Hamiltonian \cite{Mostofi2008,Marzari2012}. To calculate the integral of Berry curvature in momentum space, a dense $k$-mesh grid of $400\times400\times400$ is adopted. Fig. \ref{figure3}(a) shows the intrinsic Hall conductivity $\sigma_{xy}^{\mathrm{3D}}$ of Ba$_{2}$Cr$_{7}$O$_{14}$ relative to the Fermi level. As expected, the calculated Hall conductivity exactly match our topological analysis of parity eigenvalues. The Hall conductivity exhibits the quantized plateau ($\sigma_{xy}^{\mathrm{3D}}=-3e^2/hc$) when the Fermi level is inside the bulk band gap. Otherwise, the intrinsic Hall conductivity rapidly decreases away from the gap.

\begin{table}
\caption{Product of parity eigenvalues of occupied Bloch states at the TRIM points for Ba$_{2}$Cr$_{7}$O$_{14}$ and $h$-Fe$_3$O$_4$, respectively.}
  \begin{tabular}{p{1.7 cm}| p{1.5cm} *{1}{p{1.2cm}} *{3}{p{1.5cm}} }%{cc|cc}%{c|ccc} %p{0.8cm}|*{1}{p{2.0cm}} *{2}{p{0.75cm}
  \hline
  \hline
\centering TRIM points   & \centering  $\Gamma$  & \centering  $L$ ($\times3$)  & \centering  $F$ ($\times3$) & \ \ \ \ \ \  $Z$ \\
 \hline
\centering Ba$_{2}$Cr$_{7}$O$_{14}$  &\centering  -  & \centering  +  & \centering  + & \ \ \ \ \ \  - \\
\centering $h$-Fe$_3$O$_4$  &\centering  -  & \centering  -  & \centering  + & \ \ \ \ \ \  + \\
  \hline
  \hline
  \end{tabular}
  \label{tableI}
\end{table}

The quantized plateau of Hall conductivity in a 3D QAHE insulator will generate the corresponding topologically protected surface states. It is well known that the 2D QAHE insulator possesses chiral edge states associated with the nonzero Chern number. In a 3D QAHE insulator, the Chern number shows no dispersion along the magnetic axis. Hence, the 3D QAHE insulator will exhibit topologically protected chiral surface sheet states, which are uniformly distributed on a surface that is parallel to the magnetization direction [see Fig. \ref{figure-chial}(a)]. In contrast, the surface perpendicular to the magnetic axis will host trivial surface states. For Ba$_{2}$Cr$_{7}$O$_{14}$ with [001] magnetization, the local density of state (LDOS) is calculated using the Wannier TB Hamiltonian with the iterative Green's function method \cite{Sancho1984} as implemented in WannierTools package \cite{Wu2017wanntool}, and the LDOS projected on the semi-infinite (100) and (001) surfaces are shown in Figs. \ref{figure3}(b) and \ref{figure3}(c), respectively. As expected, the appearance of the chiral surface sheet states along $\tilde{\Gamma}-\tilde{X}$ on the (100) surface perfectly agrees with the topological prediction, indicating that the 3D QAHE is realized on this surface. In contrast, there is no topologically protected surface states on the (001) surface, showing trivial insulating features.

\begin{figure}
\setlength{\belowcaptionskip}{-0.35cm}
\setlength{\abovecaptionskip}{-0.05cm}
%	\centering
	\includegraphics[scale=0.28]{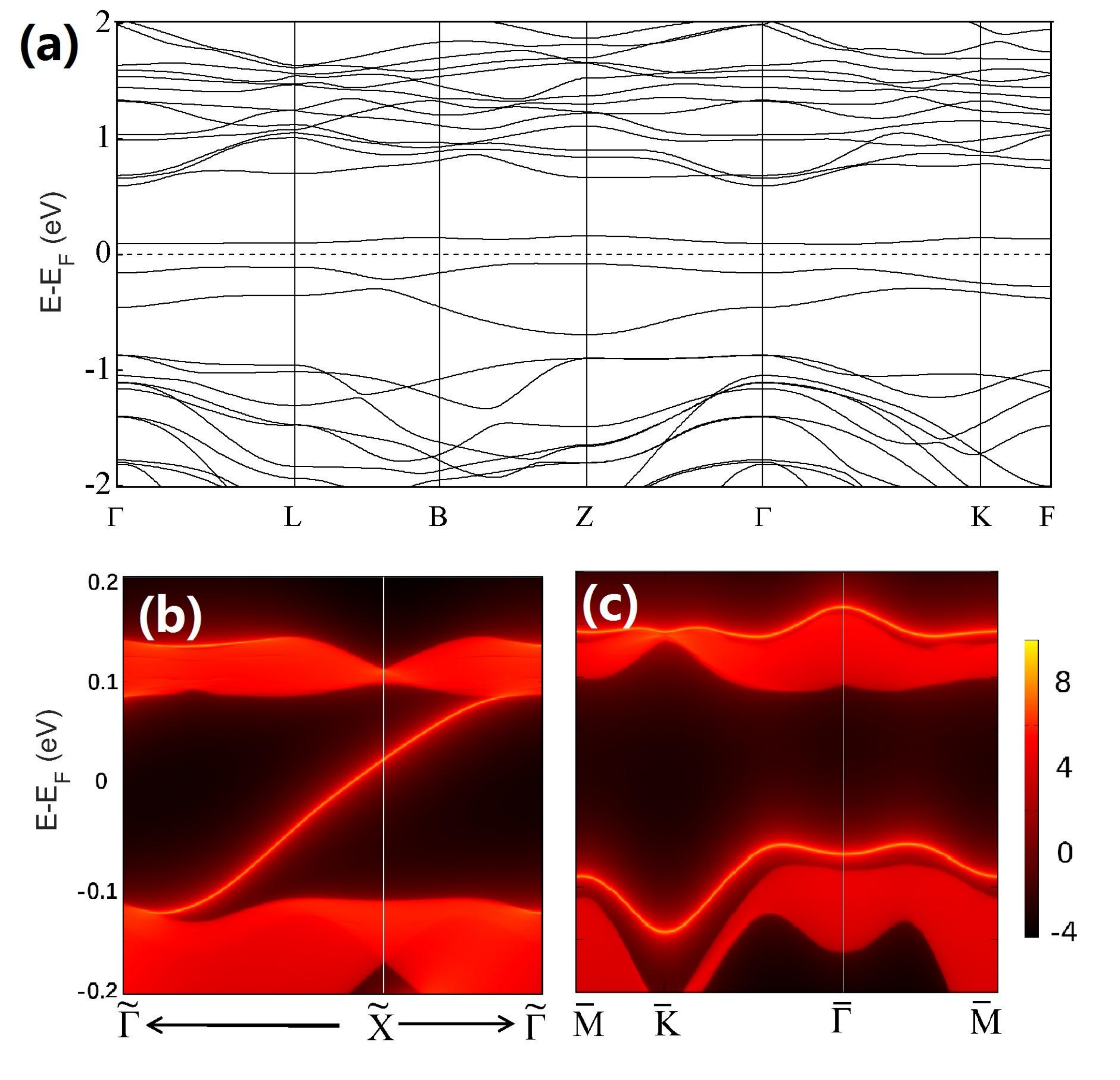}
	\caption{(a) Band structure of $h$-Fe$_3$O$_4$ with SOC along [001] magnetization. (b) The LDOS of $h$-Fe$_3$O$_4$ projected on the (100) surface with the chiral surface sheet state. (c) The LDOS of $h$-Fe$_3$O$_4$ projected on the (001) surface with the trivial surface states.
\label{figure4}}
\end{figure}

Significantly, our proposed recipe suggests that 3D QAHE can emerge in non-layered FM insulators. To verify this, we show that $h$-Fe$_3$O$_4$ crystalized in a rhombohedral structure with space group $R\bar{3}m$ (No. 166) (see SM \cite{SM}) hosts 3D QAHE. Our results show that $h$-Fe$_3$O$_4$ in a FM phase has a magnetic moment of $\sim$4.67 $\mu_B$ per Fe atom . With the magnetization along the [001] direction, the band structure of insulating $h$-Fe$_3$O$_4$ in the presence of SOC is shown Fig. \ref{figure4}(a). In comparison with Ba$_{2}$Cr$_{7}$O$_{14}$, the bulk band gap of $h$-Fe$_3$O$_4$ is considerably large, up to $\sim$ 168 meV.  From the parity products of occupied Bloch states at the TRIM points listed in Table \ref{tableI}, we can obtain that there are two paralleled planes with the first Chern number $\mathcal{C} = 1$ in reciprocal space, agreeing with the proposed recipe. Hence, we conclude that $h$-Fe$_3$O$_4$ is a 3D QAHE insulator with quantized Hall conductivity $\sigma_{xy}^{\mathrm{3D}}=-3e^2/hc$. As shown in Figs. \ref{figure4}(b) and \ref{figure4}(c), respectively, the LDOS projected on the (100) surface clearly exhibits the topologically chiral surface sheet state while the LDOS projected on the (001) surface shows the trivial surface states.

In summary, we proposed a general strategy based on symmetry considerations for exploring 3D QAHE insulators in FM materials. With the inversion symmetry, the topological constraints based on the parity analysis at the TRIM points are developed to identify the existence of 3D QAHE insulators. Our first-principles calculations show that 3D QAHE can be realized in a family of FM insulating oxides, including layered and non-layered structures, which possesses the quantized Hall conductivity. The topological features are confirmed from the chiral surface sheet states, which is uniquely distributed on the (100) surface. More importantly, the 3D QAHE would be more easily observed in experiments since the bulk materials have the higher thermodynamic stability than 2D ones.

~~~\\
~~~\\

 This work is supported by the National Natural Science Foundation of China (NSFC, Grant Nos.11674148, 11334004, and 11304403), the Guangdong Natural Science Funds for Distinguished Young Scholars (No. 2017B030306008), and the Fundamental Research Funds for the Central Universities of China (No. 106112017CDJXY300005).\\
~~~\\
Y.J.J., R.W., and B.W.X  contributed equally to this work.

%\bibliographystyle{apsrev}
%\bibliographystyle{naturemag}
%\bibliography{topological-ref}

\end{document}